\documentclass[prl,aps,groupedaddress,twocolumn,showpacs]{revtex4}

\draft

\usepackage{graphicx}
\usepackage{dcolumn}
\usepackage{amsmath}
\usepackage{bm}
\begin{document}

\title {Dephasing in quantum dot molecules via exciton-acoustic phonon coupling}
\author{G. Parascandolo}
\email{gaetano.parascandolo@epfl.ch}
\affiliation{Institute of Theoretical Physics, Ecole Polytechnique F\'ed\'erale de Lausanne (EPFL), CH-1015 Lausanne, Switzerland}
\author{V. Savona}
\affiliation{Institute of Theoretical Physics, Ecole Polytechnique F\'ed\'erale de Lausanne (EPFL), CH-1015 Lausanne, Switzerland}

\date{\today}


\begin{abstract}
We develop a theory of the linear optical spectrum of excitons in quantum dot molecules, including the effect of exciton-phonon coupling beyond the Markov limit. The model reproduces the general trend of the zero-phonon line broadening as a function of interdot distance, that were recently measured. The unexpectedly broad linewidths and their large variation are explained in terms of both the non-Markov nature of the coupling and of the matching of the phonon wavelength to the interdot distance.
\end{abstract}

\maketitle

Semiconductor quantum dots (QDs) are often considered as candidate devices for a solid-state implementation of quantum information processing \cite{Loss_PRA_98,Loss_PRB_1999,Biolatti_PRL_2000,Piermarocchi_Science_2003}. This interest is justified by the alleged long-lasting coherence of the interband polarization, which might be exploited for storing quantum information. Robustness against decoherence in QDs is usually expected as a result of quantum confinement, implying discrete energy levels below the semiconductor bandgap with a very restricted phase space available for various scattering mechanisms. This simplified view was however recently questioned, after the observation of a very efficient decoherence mechanism due to strong electron-phonon coupling \cite{Besombes01,Borri01,Bastard_PRB_03}. The strongly localized polarization in an excited QD can induce virtual inelastic phonon emission and reabsorption processes which act as a phase-destroying mechanism. Theoretically, this stems from the exact solution of a two-level system coupled to a phonon bath -- the so called {\em independent boson model} \cite{Mahan}. In practice, the decoherence rate of the interband polarization is extremely fast in the first few picoseconds following excitation and vanishes afterwards. To this corresponds a distinctive spectral feature with a sharp zero-phonon line (ZPL) and phonon-assisted broad sidebands. Both time-resolved and spectral signatures of this mechanism have been characterized by several groups \cite{Besombes01,Borri01,Bastard_PRB_03,Zimmermann_ICPS,Krummheuer02,Zimmermann_PRL_04,Langbein04}.

According to the independent boson model, the ZPL relative to a single QD electron-hole level is not broadened. The measured ZPL linewidth, and the consequent dephasing time in the nanosecond range \cite{Borri01}, are seemingly related to the radiative recombination process \cite{Langbein04}. A recent theoretical analysis \cite{Zimmermann_PRL_04} has however suggested that virtual phonon-assisted transitions to excited levels might provide a dephasing of the ZPL, with a rate comparable to the radiative one. 

Recently, the system of two vertically stacked QDs, called QD molecules, has been experimentally studied \cite{Bayer_Science_2001,Borri_PRL_2003,Bayer_PRB_2005}. A QD molecule might in principle provide the minimal system of two coupled q-bits required for the implementation of quantum gates \cite{Loss_PRB_1999,Biolatti_PRL_2000}. The electron and hole wave functions in QD molecules can tunnel through the thin barrier separating the QDs, giving rise to two bright (and two dark) exciton states \cite{Bayer_PRB_2005,Zunger_PRL_04}, whose energies are split by the effect of tunneling through the barrier and by Coulomb interaction. In a recent experimental study using coherent ultrafast four-wave-mixing spectroscopy \cite{Borri_PRL_2003}, the decoherence time of the exciton interband polarization in QD molecules has been measured. It turns out that the ZPL linewidths are generally larger than in the case of a single QD by almost one order of magnitude. Furthermore, the linewidths decrease dramatically as the interdot distance is increased. Quantitatively, these larger linewidths can only partially be explained in terms of the larger volume of exciton states in a QD molecule, while the strong dependence on the interdot distance is still unexplained. 

In this letter we present a model of exciton optical response in a QD molecule, accounting for the strong exciton-acoustic-phonon interaction. The QD exciton states are described within the effective mass scheme and the electron phonon coupling is treated within the second order Born approximation in the non-Markov limit. This limit is known to reproduce qualitatively -- and to a good extent quantitatively -- the results of the independent boson model \cite{Krummheuer02}. Linear optical spectra are computed at varying interdot distance. The measured behaviour of the ZPL linewidth is well reproduced. The larger linewidths are demonstrated to originate from the phonon-assisted coupling between phonon sidebands of one level and ZPL of a distinct level, lying at the same spectral position. This mechanism is essentially analogous to the excited-level scattering \cite{Zimmermann_PRL_04}, but much more efficient due to the spectral overlap. The strong dependence on the distance is instead attributed to a wave-matching effect between the phonon wavelength and the interdot distance, which enhances the phonon-assisted scattering from bright to dark states. 

\begin{figure}[ht]
\includegraphics[width=0.45 \textwidth]{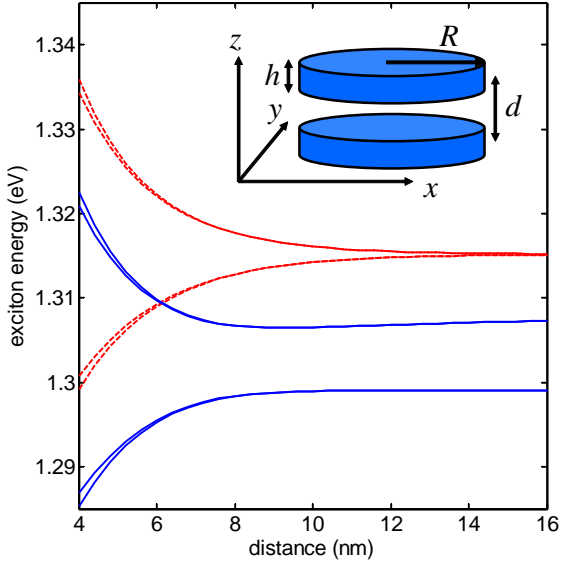}
\caption{Electron-hole pair energies (dashed) and exciton energies (solid line) as a function of the interdot distance $d$. Inset: the geometry of the QD molecule.}
\label{fig1}
\end{figure}

The dot molecule consists of two vertically stacked cylindrical QDs at mutual distance $d$, as sketched in Fig.~\ref{fig1}. We assume throughout this work that the two QDs have identical shape and size, with radius $R=10$ nm and QD height $h=2$ nm. Electron and hole states are described within two-band effective-mass approximation \cite{parascandolo05}. As material parameters for InAs QDs, we assume finite band offsets $V^e=380~\mbox{meV}$ and $V^h=200~\mbox{meV}$ between the QD and the barrier material. We take $m_e=0.067~m_0$, $m_{h\perp}=0.34~m_0$, and $m_{h||}=0.11~m_0$ for the electron, vertical and in-plane hole effective mass respectively. For the effective-mass calculations we adopt cylindrical coordinates ($\pmb{\rho},z$). In this geometry, when considering only QD-confined levels, a very good approximation consists in a separation of in-plane and $z$ variables. Given the large energy separation of higher levels, compared to the Coulomb and deformation potential interactions, we can restrict our calculations to the two lowest electron and hole levels. Without Coulomb interaction, the valence and conduction wave functions can therefore be written as $\psi_{j\gamma}({\bf r})=f_\gamma({\pmb\rho})h_{j\gamma}(z)$, with energies $E_{j\gamma}$, where $\gamma=v,c$ and $j=1,2$ for the symmetric and antisymmetric state respectively. In Fig. \ref{fig1}, both the bare electron-hole pair energies $E_{jc}-E_{lv}$ ($j,l=1,2$) as computed within the effective mass scheme, and the resulting exciton energies once Coulomb interaction (see below) is included, are plotted as a function of $d$. The energy splitting at short $d$ is mainly due to conduction electron tunneling between the two QDs and vanishes, for electron-hole pair energies, at larger $d$. The Coulomb interaction produces both a global redshift and an additional splitting of the exciton energies, this latter persisting at large values of $d$. With the present parameters, the exciton transition energies measured in Ref. \onlinecite{Borri_PRL_2003} are well reproduced. More detailed models of the electronic states \cite{Zunger_PRL_04} show that the detailed energy level structure deviates from this simple description, especially concerning the valence states. The phonon-assisted effect we are modeling, however, turns out to depend mainly on exciton-phonon coupling and, due to the lighter effective mass, almost entirely on the conduction-electron part of this coupling. The hole wave function affects only minimally our results, finally justifying our adoption of a simple effective mass scheme. 

Within this scheme, the total Hamiltonian is $\hat{H}=\hat{H}_0+\hat{H}_c+\hat{H}_{ph}^{(d)}+\hat{H}_{ph}^{(nd)}$, where $\hat{H}_0=\sum_{i,\gamma}E_{i\gamma}\hat{C}_{i\gamma}^\dagger \hat{C}_{i\gamma}$ is the bare electronic contribution in terms of the Fermi creation and annihilation operators.
We include the electron-hole Coulomb interaction in our model because the phonon-mediated effects we are describing depend on the difference in binding energies between distinct exciton levels. Neglecting intraband terms, the Coulomb Hamiltonian $\hat{H}_c$ has the usual expression
\begin{equation}
\hat{H}_c=\frac{1}{2}\sum_{ijlm}V^{ij}_{lm}\hat{C}_{ic}^\dagger \hat{C}_{jv}^\dagger \hat{C}_{mv}\hat{C}_{lc}\,,
\end{equation}
with the Coulomb matrix element given by
\begin{equation}\label{coul_matrix}
V^{ij}_{lm}=\int d{\bf r}_ed{\bf r}_h \psi_{ic}^*({\bf r}_e)\psi_{jv}^*({\bf r}_h)\frac{e^2/\epsilon_B}{\left|{\bf r}_e-{\bf r}_h\right|}\psi_{lc}({\bf r}_e)\psi_{mv}({\bf r}_h)\,,
\end{equation}
being $\epsilon_B=12.5$ the background dielectric constant for InAs.

The two QDs are coupled through electron-acoustic phonon interaction via the deformation potential interaction. The diagonal and non-diagonal parts of the electron-phonon interaction are
\begin{eqnarray}
\hat{H}_{ph}^{(d)}&=&\sum_{\bf q}\hbar\omega_{\bf q}\hat{a}_{\bf q}^\dagger \hat{a}_{\bf q}+\sum_{i,\gamma,{\bf q}} M_{\bf q}^{i\gamma,i\gamma}\left(\hat{a}_{\bf q}^\dagger+\hat{a}_{-\bf{q}}\right)\hat{C}_{i\gamma}^\dagger \hat{C}_{i\gamma}\,,\label{Hphd}\\
\hat{H}_{ph}^{(nd)}&=&\sum_{i\neq j,\gamma}\sum_{\bf q}M_{\bf q}^{i\gamma,j\gamma}\left(\hat{a}_{\bf q}^\dagger+\hat{a}_{-{\bf q}}\right)\hat{C}_{i\gamma}^\dagger \hat{C}_{j\gamma}\,,\label{Hphnd}
\end{eqnarray}
where the matrix element for deformation potential coupling with acoustic phonons of dispersion $\omega_{{\bf q}}=qs$ is given by
\begin{equation}\label{M_q}
M_{\bf q}^{i\gamma,j\gamma}=D_{\gamma}\sqrt{\frac{\hbar\omega_{\bf q}}{2\rho_m Vs^2}}\int{d{\bf r}\psi_{i\gamma}^*({\bf r})\psi_{j\gamma}({\bf r})e^{i{\bf q\cdot r}}}\,,
\end{equation}
being $s$ the sound velocity, $\rho_m$ the mass density, $V$ the normalization volume and $D_{\gamma}$ the deformation potential constants. The diagonal term $\hat{H}_{ph}^{(d)}$ describes the coupling of each isolated electron or hole state with the phonon modes \cite{Krummheuer02,Zimmermann_ICPS,Bastard_PRB_03}. The off-diagonal term $\hat{H}_{ph}^{(nd)}$ accounts for the phonon-assisted scattering between different electron or hole states. For our calculations, we take $s=4557$ m/s, $\rho_m=5.67$ g/cm$^3$, $D_c=-13.6$ eV and $D_v=-7.1$ eV \cite{Zimmermann_PRL_04}. 

\begin{figure}[h!]
\includegraphics[width=0.45 \textwidth]{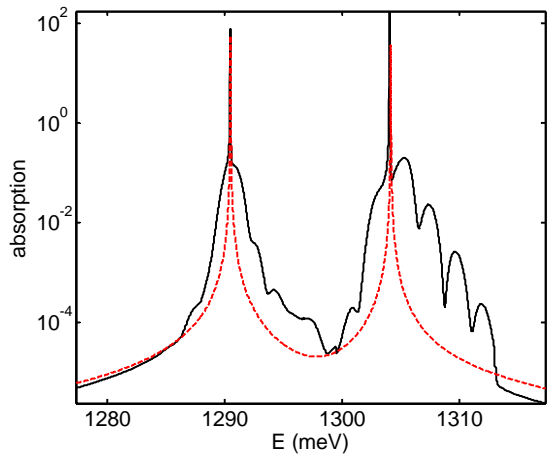}
\caption{Comparison between the non-Markov (solid) and Markov (dot-dashed) computed exciton absorption spectrum at fixed temperature $T=10$~K and interdot distance $d=6$~nm.}
\label{fig2}
\end{figure}

The independent boson model can be diagonalized exactly in the case of a single electron-hole pair state \cite{Mahan}. Here however, because of the off-diagonal coupling $\hat{H}_{ph}^{(nd)}$ and of the Coulomb interaction, the full diagonalization of the problem is a cumbersome task. Our approach therefore consists in first diagonalizing the electron Hamiltonian $\hat{H}_0+\hat{H}_c$, giving rise to exciton states $\Psi_j({\bf r}_e,{\bf r}_h)=\sum_{lm}\alpha_j^{lm}\psi_{lc}({\bf r}_e)\psi_{mv}({\bf r}_h)$, with energy $E_j$ ($j=1,\ldots,4$). The exciton-phonon Hamiltonians (\ref{Hphd}) and (\ref{Hphnd}) can be consistently rewritten in this basis in terms of exciton operators $\hat{B}_j,\hat{B}^\dagger_j$. We then develop a density matrix formalism for the linear interband polarization, in the framework of the second-order Born approximation for the exciton-phonon interaction, without performing the Markov limit. It is known that this approach can quantitatively fairly well reproduce the exact result in the single QD case~\cite{Krummheuer02}. For comparison, we will evaluate the Markov limit as well. The end equation for the four excitonic polarizations $\Pi_j(\omega)=\langle\hat{B}^\dagger_j(\omega)\hat{B}_j(t=0)\rangle$ reads
\begin{equation}\label{polarizations}
\left[E_j-\hbar\left(\omega+i\gamma_j\right)\right]\Pi_j(\omega)-\sum_m\Sigma_{jm}(\omega)\Pi_m(\omega)=g_j\,,
\end{equation}
where the exciton-phonon self energy is given by $\Sigma_{jm}(\omega)=\sum_{k,{\bf q}}M_{\bf q}^{jk}D_{\bf q}^{(k)}(\omega)M_{-\bf q}^{km}$, being $M_{\bf q}^{jk}$ linear combinations of $M_{\bf q}^{i\gamma,j\gamma}$ in the Coulomb basis, and $D_{\bf q}^{(k)}(\omega)$ are the phonon propagators~\cite{Mahan} (we assume a thermal phonon distribution at the lattice temperature). Here, $\hbar\gamma_j$ is the radiative linewidth of the $j$-th excitonic state, computed as in Ref.~\cite{Kavokin_APL_2002}, and $g_j=|\int\Psi_j({\bf r},{\bf r})d{\bf r}|$ its oscillator strength. We numerically solve Eq.~(\ref{polarizations}) for the exciton polarizations $\Pi_j(\omega)$. The optical spectrum is then related to the imaginary part of these quantities. A typical exciton spectrum for $d=6$ nm is plotted in Fig.~\ref{fig2}. Two of the four exciton wave functions are even with respect to $z$-inversion and the corresponding transitions are therefore optically active, giving rise to the two peaks in the calculated spectrum \cite{Bayer_PRB_2005,Zunger_PRL_04}. Each peak is characterized by a narrow zero-phonon line (ZPL) and by phonon sidebands originating from the diagonal elements $\Sigma_{ii}(\omega)$ of the exciton-phonon self-energy in (\ref{polarizations}), as in the single QD case~\cite{Krummheuer02,Zimmermann_ICPS}. The two other exciton states are dark. All four exciton states, however, are coupled to each other via the off-diagonal terms of the exciton-phonon self-energy in (\ref{polarizations}). The oscillations of the sidebands are an artefact of the Born approximation and are known to be absent in the exact solution of the independent boson model~\cite{Krummheuer02}. For comparison, we also plot the lorentzian spectrum computed in the Markov limit. 

\begin{figure}[t]
\includegraphics[width=0.45 \textwidth]{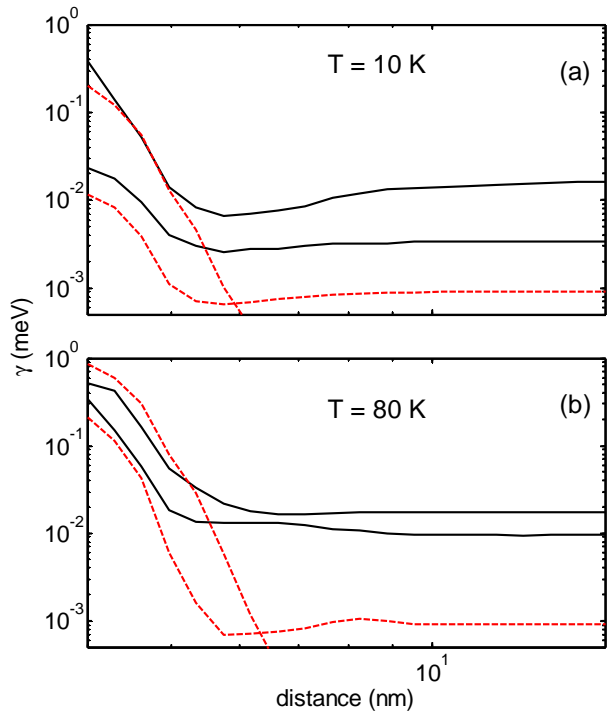}
\caption{(a) Computed ZPL linewidths as a function of the interdot distance $d$ at $T=10$ K. Solid line: non-Markov. Dashed lines: Markov limit. (b) Same as in (a) but for $T=80$ K.}
\label{fig3}
\end{figure}

Fig. \ref{fig3}(a) shows the computed ZPL linewidths of the optically active excitons as a function of $d$, together with the corresponding result in the Markov limit. The exciton-phonon coupling has two unexpected consequences. The first is a large broadening of both ZPLs at short QD distance $d$, which quickly decreases as $d$ is increased. The same trend was measured by Borri {\em et al}, and left unexplained. From our calculation we can infer that this linewidth enhancement is due to transitions between bright and dark exciton states via phonon emission or absorption. Let us consider only the conduction electron part of the exciton state, as the phonon coupling is much less effective for the valence band due to the heavy mass. The two conduction states have opposite parity with respect to spatial $z$-inversion. In order for the integral in Eq. (\ref{M_q}) to be large, therefore, the phonon wavelength has to be an integer multiple of the interdot distance, thus restoring the even parity in the argument of the integral. If this condition is satisfied together with energy conservation, the scattering process is enhanced. For $d=4$ nm, the bright-dark energy splitting is $1.6$ meV which, according to the phonon dispersion, corresponds to a phonon wavelength $\lambda\simeq12$ nm, exactly three times the interdot distance. This wave-matching condition is quickly lost for larger $d$, resulting in the dramatic linewidth decrease. This effect would be equally predicted by a standard Boltzmann dynamics and therefore appears also in the Markov limit. The second consequence of exciton-phonon coupling observed in the data of Fig. \ref{fig3}(a) are the large values of the ZPL linewidths persisting in the non-Markov case even at large interdot distance. A similar behaviour was observed in the experiment by Borri {\em et al} \cite{Borri_PRL_2003}. The computed linewidths must be compared to the Markov result which, for large $d$, gives much smaller values. The Markov values at large $d$ are essentially given by the computed radiative linewidths, which are very different for the two bright exciton states due to their different symmetry \cite{Bayer_PRB_2005,Zunger_PRL_04}. The large measured linewidths are therefore explained as being a consequence of the non-Markov nature of the coupling and of the coupled dynamics described by Eq. (\ref{polarizations}). Due to the close energy spacing, each ZPL is spectrally superposed to the tail of the phonon sideband relative to the other exciton levels, including the dark ones. Via the off-diagonal elements of the self-energy in (\ref{polarizations}), the ZPL is thus coupled to a continuum of levels, resulting in a spectral line broadening. This mechanism is analogous to the broadening of the ZPL due to virtual phonon-assisted transitions to excited QD levels, recently described by Muljarov {\em et al} \cite{Zimmermann_PRL_04}, which can account for the ZPL broadening measured for single QDs \cite{Borri01}. Due to the close energy spacing of the four exciton levels, however, we predict that this is the most effective line-broadening mechanism for ZPLs in a QD molecule. At low temperature, phonon sidebands are strongly asymmetric, with a much larger high-energy side due to phonon emission as compared to phonon absorption probability. Therefore, the linewidth of the high-energy ZPL, according to the present mechanism, is expected to be larger than the low-energy one. This is seen in Fig. \ref{fig3}(a), where the two ZPL linewidths differ by approximately an order of magnitude in the non-Markov case. The same difference was experimentally observed~\cite{Borri_PRL_2003} and attributed to the larger probability of relaxation from the upper to the lower level through phonon emission. At higher temperature, the phonon sidebands become more symmetric and, as a consequence, our mechanism predicts similar ZPL linewidths for the two spectral peaks, as shown in Fig. \ref{fig3}(b) where the computed ZPL linewidths for T=80 K are plotted. This further supports our interpretation. We expect however that, for this range of temperatures, coupling to other states via both acoustic and optical phonons come into play, thus requiring an extension of the present model.

In conclusion, we modeled the optical response of excitons in a QD molecule, including the coupling to acoustic phonons beyond the Markov limit. Two new effects, absent in the single QD case, dominate here the ZPL broadening. The first is the phonon scattering to dark states which is enhanced when the phonon wavelength matches the interdot distance. The second is the off-diagonal phonon-assisted coupling which, in the non-Markov limit, produces a spectral broadening of the ZPLs significantly larger than the radiative linewidth. All experimental findings can be explained in the the context of the present model. These broadening mechanisms provide, for QD molecules, a principial upper bound to the characteristic dephasing times, and represent therefore an even more restricting limitation, than in the single QD case, to the applications in quantum information technology.

We acknowledge financial support from the Swiss National Foundation through project
N. 620-066060.

\bibliographystyle{PRSTY}

\end{document}